\documentclass[usenatbib]{mn2e}

\newcommand{\mnras}{MNRAS}
\usepackage{color}
\usepackage{graphicx}
\usepackage{perpage} 
\usepackage{pdflscape}
\MakePerPage{footnote} 
\definecolor{grey}{rgb}{0.5,0.5,0.5}

\title[Optical polarization map of the Polaris Flare]{Optical polarization map of the Polaris Flare with RoboPol}
\author[G. Panopoulou et al.]
{G. Panopoulou$^{1,2}$\thanks{Contact authors' e-mail addresses:
panopg@physics.uoc.gr (GP); tassis@physics.uoc.gr (KT)},
K. Tassis$^{1,2 \star}$,  D. Blinov$^{1,7}$, V. Pavlidou$^{1,2}$,
O.\, G. King$^3$, 
\newauthor
E. Paleologou$^1$,  A. Ramaprakash$^4$,
 E. Angelakis$^{5}$,
 M. Balokovi\'{c}$^3$, 
 H. K. Das$^4$,  
\newauthor
R. Feiler$^6$, T. Hovatta$^{3,8}$, P. Khodade$^4$, S. Kiehlmann$^5$, A. Kus$^6$, N. Kylafis$^{1,2}$,
\newauthor
   I. Liodakis$^{1}$, D. Modi$^{4}$, I. Myserlis$^{5}$, 
  I. Papadakis$^{1,2}$, I. Papamastorakis$^{1,2}$,
\newauthor 
 B. Pazderska$^6$, E. Pazderski$^6$, T.\,J. Pearson$^{3}$, C. Rajarshi$^4$,  A.\,C.\,S. Readhead$^{3}$,
\newauthor
 P. Reig$^{2,1}$, J.\,A. Zensus$^5$\\
$^{1}$Department of Physics and ITCP\thanks{Institute for
  Theoretical and Computational Physics, formerly Institute for Plasma
Physics}, University of Crete, 71003, Heraklion, Greece\\
$^{2}$Foundation for Research and Technology - Hellas, IESL, Voutes, 7110 Heraklion, Greece\\
$^{3}$Cahill Center for Astronomy and Astrophysics, California Institute of Technology, 1200 E California Blvd, MC 249-17,\\Pasadena CA, 91125, USA\\
$^4$Inter-University Centre for Astronomy and Astrophysics, Post Bag
4, Ganeshkhind, Pune - 411 007, India\\
$^{5}$Max-Planck-Institut f\"{u}r Radioastronomie, Auf dem H\"{u}gel
69, 53121 Bonn, Germany\\
$^6$Toru\'{n} Centre for Astronomy, Nicolaus Copernicus University, Faculty of Physics, Astronomy and Informatics,\\
Grudziadzka 5, 87-100 Toru\'{n}, Poland\\
$^7$Astronomical Institute, St. Petersburg State
University,Universitetsky pr. 28, Petrodvoretz, 198504 St. Petersburg,
Russia\\
$^8$Aalto University Mets\"ahovi Radio Observatory, Mets\"ahovintie 114, 02540 Kylm\"al\"a, Finland
}

\begin{document}


\maketitle

\label{firstpage}

\begin{abstract} 
The stages before the formation of stars in molecular clouds are poorly understood. 
Insights can be gained by studying the properties of quiescent clouds, such as their
magnetic field structure. The plane-of-the-sky orientation of the field can be traced by 
polarized starlight.
We present the first extended, wide-field ($\sim$10 $\rm deg^2$) map of the Polaris Flare
cloud in dust-absorption induced optical polarization of background stars, using the 
RoboPol polarimeter at the Skinakas Observatory. This is the first application of the
wide-field imaging capabilities of RoboPol. The data were taken in the R-band and
analysed with the automated reduction pipeline of the instrument. We present in detail optimizations in the reduction pipeline specific to wide-field observations.
Our analysis resulted in reliable measurements of 648 stars with median 
fractional linear polarization 1.3\%.
The projected magnetic field shows a large scale ordered pattern. At high longitudes 
it appears to align with faint striations seen in dust emission, while in the central 4-5 deg$^2$
it shows an eddy-like feature. The overall polarization pattern we obtain is in good agreement
with large scale measurements by Planck of the dust emission polarization in the same area of the sky.
\end{abstract}


\section{Introduction}
\label{intro}

Molecular clouds in their vast complexity hold the key to understanding the early stages of the 
star formation process. Magnetic fields and turbulence are the two main mechanisms that dictate the structural, 
dynamical and evolutionary properties of these clouds, through their competition against gravity. 
Their role in the onset of star formation can be studied best in quiescent non--star-forming regions, where 
stellar feedback is not present. One such region is the Polaris Flare, a translucent molecular cloud at a distance
between 130 pc and 240 pc \citep{heithausen}, discovered by \cite{Heiles}. 
It is believed to be in the early stages of its formation, 
since it does not exhibit any signs of active star formation \citep{andre}. CO observations have provided invaluable 
information on the turbulence signatures in the densest parts of the cloud \citep{falgarone1998,hily-blant}. 
Recently, the Herschel space telescope mapped over 15 deg$^2$ of the cloud in dust emission \citep{andre,miville}. 

The structure of the magnetic field of a cloud, as projected on the plane of the sky,
can be probed by observing polarized radiation.
The polarization of starlight transmitted through a cloud is believed to be caused by dichroic extinction 
due to aspherical dust grains that are partially aligned with the magnetic field. This alignment causes the 
polarization direction of the light of background stars to trace the magnetic field direction of the cloud as 
projected on the plane of the sky. 
The same alignment process causes the thermal emission of these dust grains to be polarized in the direction 
perpendicular to the magnetic field. 

Information on the plane-of-the-sky magnetic field of the Polaris Flare has been provided for the first time 
through polarized dust emission \citep{planckXIX,planckXX}. 
These data, however, are limited by the instrumental 
resolution and confusion along the line of sight. A mapping of the region in polarized starlight, which is 
complementary to the dust emission but suffers from different limitations, is necessary to resolve these issues.

We obtained optical polarization measurements of stars projected on 10 deg$^2$ of the Polaris Flare region 
with RoboPol.
The RoboPol instrument is a 4-channel optical polarimeter with 
no moving parts, other than a filter wheel (Ramaprakash et al., in prep). 
It can measure both linear fractional Stokes parameters $q=Q/I$ and $u=U/I$ simultaneously, 
thus avoiding errors caused by the imperfect alignment of rotating optical elements
and sky changes between measurements (polarization, seeing conditions).

\begin{figure}
\centering
\includegraphics[scale=1]{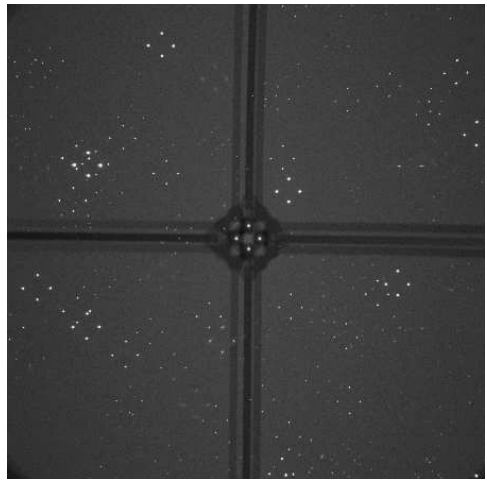}
\caption{A field observed by RoboPol. Each star in the field creates a quadruplet of images (spots) on the CCD. 
The central dark region is the mask used for lowering sky noise for the target at the centre of the field
of view and the cross-like figure is created by the mask-supporting legs.}
\label{fig:typicalfield}
\end{figure}

Each star in the field of view creates four images (spots) on the CCD displaced symmetrically 
in the horizontal and vertical directions.
A mask supported by four legs is positioned at the centre of the field of view. This allows targets
that are centred on the mask to be measured with four times lower sky noise than the rest of the field.
A typical image seen with RoboPol is shown in Fig. \ref{fig:typicalfield}. 
The instrument has a $13'\times13'$ field of view, enabling the rapid polarimetric mapping of large areas 
of the sky.
RoboPol is equipped with standard Johnson-Cousins $R$- and $I$-band filters
and is mounted on the 1.3-m, f/7.7 Ritchey--Cr\'etien telescope at Skinakas Observatory in Crete, Greece. 
It has been operating since May 2013. 

RoboPol has been monitoring the optical linear polarization of a large sample of gamma-ray
bright blazars for the past two years \citep{2014MNRAS.442.1693P}. 
In addition, the instrument is being used for long-term monitoring of Be X-ray binaries
\citep{2014MNRAS.445.4235R}. 
Observations of optical afterglows of gamma-ray bursts have also been conducted with RoboPol
\citep{2014MNRAS.445L.114K}. More complete descriptions of the instrument and data reduction pipeline
are given in Ramaprakash et al. (in prep.) and \cite{2014MNRAS.442.1706K}, respectively.

The data presented here are the first obtained from an analysis of the instrument's entire field of view. 
We present the observational details in Section \ref{sec:observations}. In Section
\ref{sec:pipelinecuts} we describe the methods used for analysing sources in the entire field of view.
We present and discuss the results of our observations in Section \ref{sec:PF} and
summarize our findings in Section \ref{sec:summary}.

\section{Observations}
\label{sec:observations}
Observations were taken during 25 nights from August to November 2013, totaling around 60 hours of 
telescope time.
The observations covered an area of 10 deg$^2$: $l =$ [122.6$^\circ$, 126.0$^\circ$], $b =$ [24.7$^\circ$, 27.9$^\circ$]. 
The area was initially divided into 275 non-overlapping fields spaced 13.2 arcminutes apart (slightly larger than the size of the RoboPol field of view). Of them, 227 were observed by the end of the period. The number of observations of each field ranges between 2 and 6, with 93\% of all fields having been observed at least 3 times. The exposures were either 120 or 180 seconds long, with 95\% of the exposures having lasted 180 s. All observations were taken in the R band.

\section{Analysis}
\label{sec:pipelinecuts}

Previous studies with RoboPol concentrated on sources either exclusively within the mask, or 
with the addition of some selected sources in the field of view around the central target.
Although the data reduction pipeline presented by \cite{2014MNRAS.442.1706K} was designed for the
entire RoboPol field of view, its implementation in this particular project showed the need for
some adjustments and additions. Sources outside the mask present a number of challenges. Some are common in most 
polarimetric studies in the optical, while others are due to the particular design of the instrument.
A measurement may be adversely affected by one of the following sources of systematic error: 
\begin{itemize}
\item Large-scale optical aberrations
\item Proximity to the mask and its legs
\item Proximity to the CCD edge
\item Proximity to other sources
\item Selection of apertures for photometry
\item Dust on optical elements
\end{itemize}

An additional systematic error has already been identified and discussed by 
\cite{2014MNRAS.442.1706K}. A rotation in the polarization reference frame of the telescope with respect to
that of the sky causes all angles to be larger by $2.31^\circ\pm 0.34^\circ$.  
All polarization angle measurements presented in this paper have been corrected for this.

This section outlines the analysis of observations and the methodology adopted to control these systematic effects.

\subsection{Significance of measurements and debiasing}
The measurement of the fractional linear polarization ($p$) at the low polarization regime which is
relevant for interstellar polarization, being always positive, is biased towards values larger than the 
true (intrinsic) polarization \citep{simons-stewart}. 
Thus, $p$ measurements should be debiased to find their most probable intrinsic value. 
In the analysis, we consider only sources with signal-to-noise ratios 
($p/\sigma_p \geq 2.5$) so that errors are approximately normally distributed.
The maximum-likelihood estimator of the true value of $p$ found by \cite{vaillancourt} 
for measurements with $p/\sigma_p \geq 3$ significance is:
\begin{equation}
\centering
p_d = \sqrt{p^2-\sigma_p^2}
\label{eqn:debiasing}
\end{equation} 
We extend this formula to $p/\sigma_p \geq 2.5$ and use it to debias all measurements of $p$.

\subsection{Large-scale optical aberrations}
\label{ssec:model}

\begin{figure}
\centering
\includegraphics[scale=0.48]{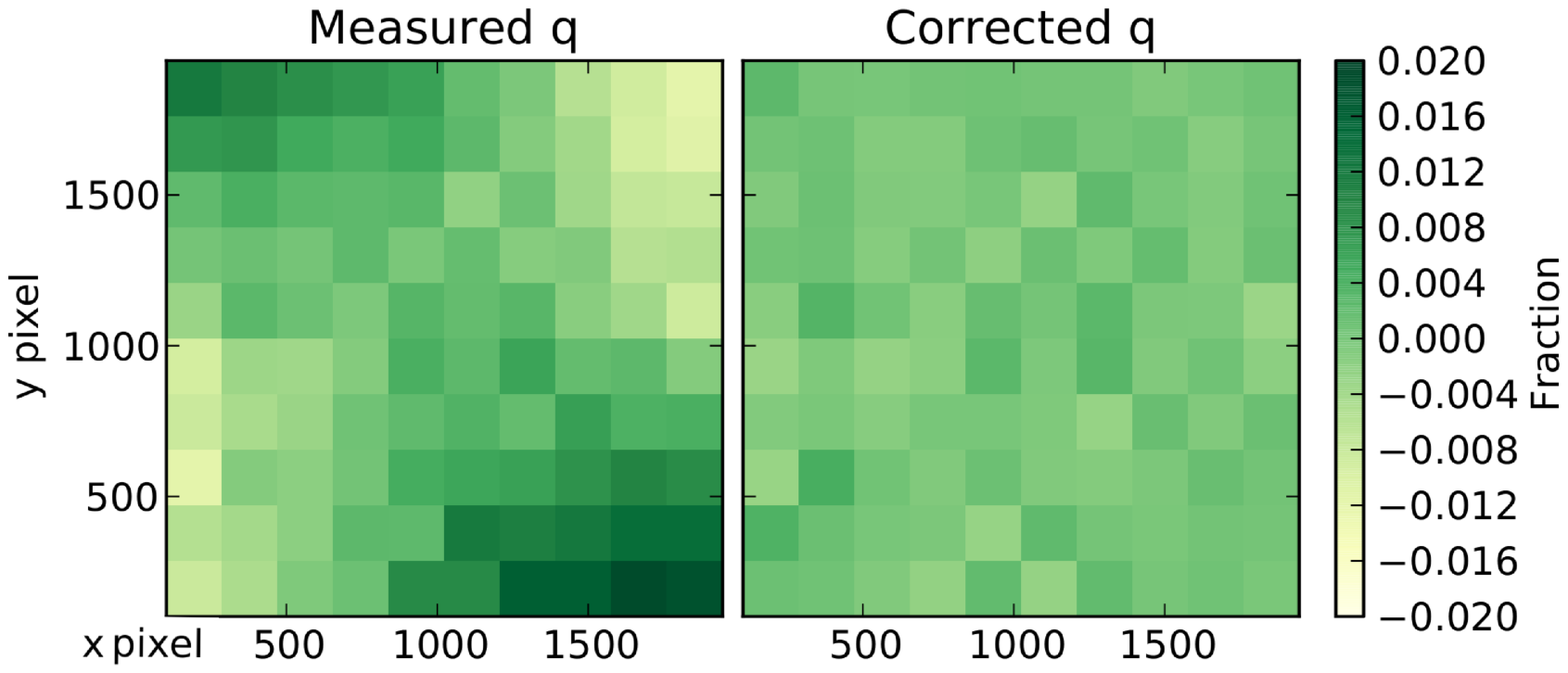}
\end{figure}
\begin{figure}
\centering
\includegraphics[scale=0.48]{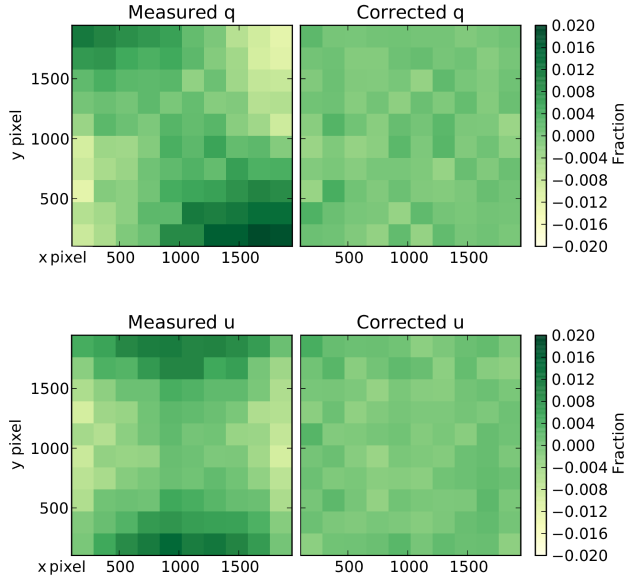}
\caption{Relative Stokes parameters across the CCD. Left: measured $q$ (top) and $u$ (bottom). Right: residual
$q$ (top) and $u$ (bottom) after subtracting the fitted model.
Each square panel shows values on the CCD binned in 100 cells. 
Each cell is colored according to its average value.}
\label{fig:modelu}
\end{figure}

Large-scale aberrations caused by the optical system are corrected by the instrument model, 
as presented by \cite{2014MNRAS.445L.114K}. The model is created by placing an unpolarized standard star
at many positions across the field of view and finding the best-fitting parameters that cancel the global, instrumentally-induced polarization and vignetting. 

The instrument model has been found to perform equally well, regardless of telescope pointing position 
(which may result in different telescope stresses) and after multiple removals and re-installations of 
the instrument on the telescope. The set of models that were created for these tests have been combined
into one with improved performance with respect to that presented by \cite{2014MNRAS.445L.114K}. Below we 
estimate the systematic uncertainty that remains after the model correction.

\subsubsection{Systematic uncertainty from model residuals}

\begin{figure}
\centering
\includegraphics[scale=1]{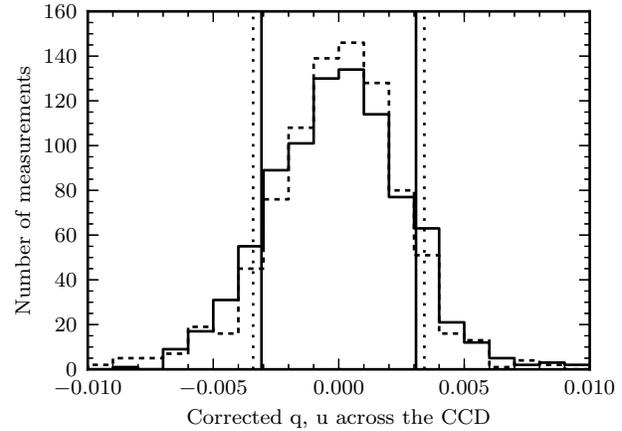}
\caption{Distributions of the residuals of $q$ and $u$ across the CCD, after the subtraction of the model fit
(q: dotted, u: solid).
The vertical lines show the standard deviation of each distribution.}
\label{fig:residual-dist}
\end{figure}

Fig. \ref{fig:modelu} shows the uncorrected (left) and corrected (right) $q$ (top) and $u$ (bottom)
values across the CCD derived by using this combined model. 
The data are binned in 100 cells of 39 arcsec width and the mean value is plotted in each one. 
On average, 2.4 star measurements contribute to each cell. The residuals appear to be homogeneous across the CCD.

The distributions of residual $q$ and $u$ of the combined model are shown in Fig. \ref{fig:residual-dist}.
Vertical lines show the standard deviation of each distribution ($\sigma_{\it q,\rm res}$ = 0.0034, 
$\sigma_{\it u,\rm res}$ = 0.0031).
Statistical errors of measurements of unpolarized standards are an order of magnitude lower 
than these standard deviations, thus their contribution to this scatter can be ignored.
Therefore, we take the systematic uncertainties in $q$ and $u$ to be 
$\sigma_{q,\rm sys} = \sigma_{\it q,\rm res},\, \sigma_{u,\rm sys} = \sigma_{\it u,\rm res}$.

From now on, in order to estimate total uncertainties in $q$ and $u$, we add statistical and systematic
uncertainties in quadrature,
\begin{equation}
\centering
\sigma_{q}^2 = \sigma_{q,\rm sys}^2 + \sigma_{q, \rm stat}^2
\label{eqn:quadratureq}
\end{equation}
\begin{equation}
\centering
\sigma_{u}^2 = \sigma_{u,\rm sys}^2 + \sigma_{u, \rm stat}^2.
\label{eqn:quadratureu}
\end{equation} 
 
The total uncertainty in $q$ and $u$ can be propagated to find the total uncertainty in 
fractional linear polarization ($p$) and electric vector position angles (EVPA or $\chi$) using the equations:
\begin{equation}
\centering
p = \sqrt{q^2 + u^2}, \, \sigma_p = \sqrt{\frac{q^2\sigma_q^2 + u^2\sigma_u^2}{q^2+u^2}}
\label{eqn:P}
\end{equation}
\begin{equation}
\chi = \frac{1}{2} \tan ^{-1}{\frac{u}{q}}, \, \sigma_{\chi} = \frac{1}{2} \sqrt{\frac{u^2\sigma_q^2 + q^2\sigma_u^2}{(q^2+u^2)^2}}
\label{eqn:EVPA}
\end{equation}

\begin{table*}
\centering
\caption{Polarization standard stars shown in Fig. \ref{fig:polstands}}
\begin{tabular}{|c|c|c|c|c|}
\hline
                & BD+59.389       & VICyg12 & HD151406 & HD212311\\
\hline
P (\%)          & 6.430 $\pm0.022$ & 7.893 $\pm0.037$ & 0.085 $\pm0.041$    &0.034 $\pm0.021$\\
				& 6.43 	$\pm0.13$ & 7.18 $\pm0.04$   &					 & 0.02$\pm0.021$\\
				&                 &                  &                     &0.045\\ 
$\chi$($^\circ$)& 98.14$^\circ$ $\pm 0.10$& 116.23$^\circ$ $\pm0.14$& -2$^\circ$ &50.99$^\circ$\\
				& 96.0 $^\circ$ $\pm 0.6$& 117$^\circ$ $\pm1$     &       & 36.2$^\circ$ $\pm51.3^\circ$\\
				&                 &                             &         & 10.4$^\circ$\\
Band            &  R              & R                           &      no filter      & V \\
Reference       & 1 ,5              & 1, 3                      & 2               & 1,4,5\\
\hline
\end{tabular}
\\1 - \cite{Schmidt}, 2 - \cite{berdyugin}, 3 - \cite{bailey}, 4- \cite{heiles2000}, 5 - \cite{eswaraiah2011}
\label{tab:refs}
\end{table*}

Assuming low polarization the expression for $\sigma_{\chi,\rm sys}$ can be written as:
\begin{equation}
\centering
 \sigma_{\chi} \simeq \frac{1}{2} \frac{\sigma_p}{p}.
\label{eqn:chi}
\end{equation}

\begin{figure}
\centering
\includegraphics[scale=1]{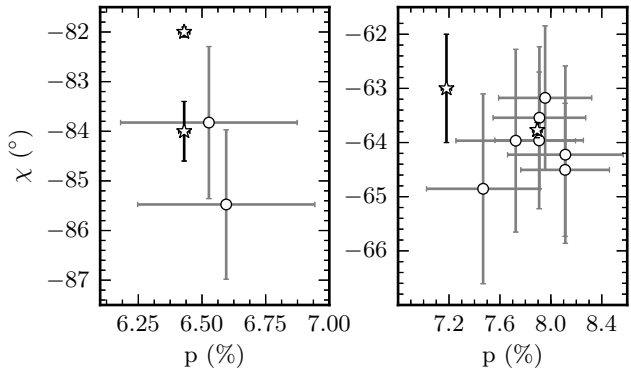}
\caption{EVPA versus fractional linear polarization of standard stars. Left: BD+59.389, right: VI Cyg12. 
Literature values are shown by stars (see references in Table \ref{tab:refs}) and circles are measurements outside the mask. Error bars include the statistical and systematic uncertainties added in quadrature.}
\label{fig:polstands}
\end{figure}

\begin{figure}
\centering
\includegraphics[scale=1]{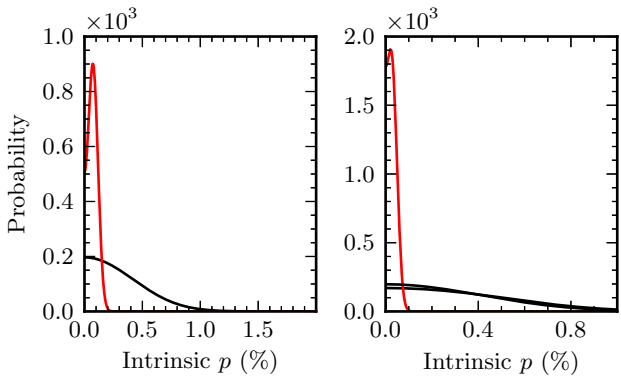}
\caption{Probability distribution of the intrinsic value of $p$ of unpolarized standards, given the measurement 
in the literature (red) and our own (black). Left: HD151406,right: HD212311. There are two black lines
(measurements) of HD212311.}
\label{fig:rice}
\end{figure}

\subsubsection{Performance: Standard Stars}
\label{ssec:calibration}
To assess the accuracy of the instrument model, measurements of stars with known polarization values 
were taken and were then compared to the literature. During the two observing seasons, a number of standard stars
(different from the ones used for the model calculation) were observed throughout the field of view. 
Catalog measurements as well as the band in which they were taken are shown in Table \ref{tab:refs}. 
Measurements of the unpolarized stars in the $R$ band could not be found, so those in other bands are quoted.

Fig. \ref{fig:polstands} presents RoboPol measurements of polarized standards
(denoted by circles) and their literature values (stars) on the EVPA - polarization fraction plane.
 All $p$ measurements are consistent with the literature within the errors (which include both the statistical and systematic uncertainties discussed above). Measurements of $p$ have not been debiased.

In the case of unpolarized stars, biasing of $p$ measurements is very pronounced and the interpretation of the 
measurement uncertainty is not straightforward. To facilitate comparison of our measurements with literature
values, we plot, in  Fig. \ref{fig:rice}, the probability distribution (likelihood) of the intrinsic (true) 
value of $p$, given the literature measurement (red) and our own (black). This likelihood function (see 
\cite{vaillancourt}, equation 8) takes into account that the measured values of $p$ follow a Rice, rather than a 
normal, distribution. In calculating the likelihood function we have used a total uncertainty obtained by adding 
statistical and systematic uncertainties in quadrature, as in equations (\ref{eqn:quadratureq}) and (\ref{eqn:quadratureu}). In both cases, our measurements are consistent within 
uncertainties with the literature 
measurements. There are two measurements (black lines) of the standard HD212311. 
For unpolarized standards the EVPA does not carry meaningful information, as can be seen by 
substituting $\sigma_p/p > 1$ into  equation (\ref{eqn:EVPA}): $\sigma_\chi \geq 30^\circ$.

\subsection{Proximity to the mask, legs and CCD edge}

The mask and its supporting legs cast shadows on specific regions of the CCD 
rendering them unusable.
Therefore, sources that happen to fall in the shadow of the mask legs or within 155 pixels (radially) 
of the mask centre are not considered in the analysis. 

Sources falling very close to any of the CCD edges are very likely to suffer partial photon losses.
Also, light reaching these areas is subject to large optical distortions.
Since the typical separation between a pair of the four images is 100 pixels, we reject any spot within 
100 pixels of the edges from the analysis.

\subsection{Proximity to other sources}

\begin{figure}
\centering
\includegraphics[scale=1]{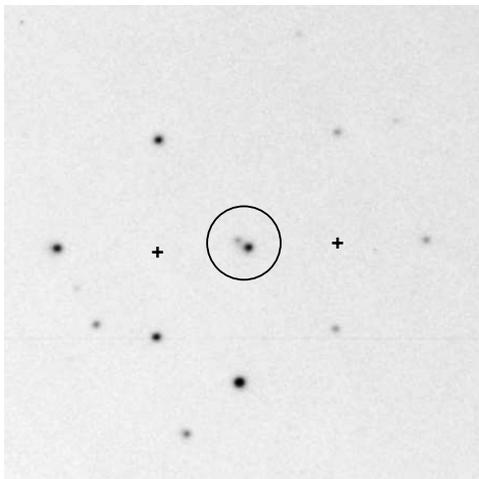}
\caption{An example of sources that are affecting each other's measurements due to their proximity
(circled spots). The positions of the stars (centres of quadruplets) are shown with crosses.}
\label{fig:spot-on-spot}
\end{figure}

\begin{figure*}
\centering
\includegraphics[scale=1]{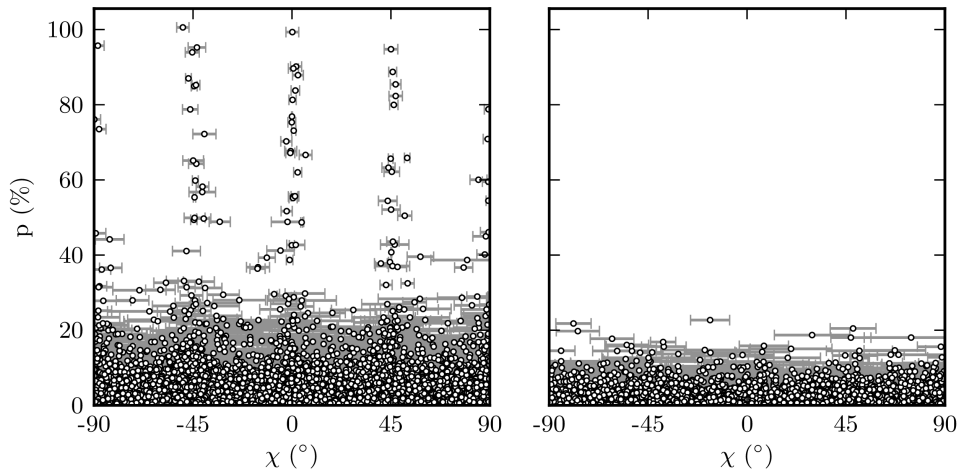}
\caption{Fractional linear polarization versus EVPA for stars in the Polaris Flare field. Left: Measurements at regular
angles (0$^{\circ}$, $\pm$45$^{\circ}$, $\pm$90$^{\circ}$) are caused by nearby contamination as seen in Fig. 
\ref{fig:spot-on-spot}. Right: Measurements that survive after the removal of stars that suffered this contamination. Most remaining measurements of $\rm p > 5\%$ are caused by other systematics.}
\label{fig:fingers}
\end{figure*}

Sometimes images from different stars happen to fall within a few pixels of each other on the CCD. 
Since the typical diameter of a spot is 8 pixels (3.2 arcsec), photons from both spots are blended, as shown in Fig. \ref{fig:spot-on-spot}. The relative Stokes parameters are computed using the following equations:
\begin{equation}
\centering
q  = \frac{N_{1} - N_{0}}{N_{1} + N_{0}},\, \sigma_q = \sqrt{\frac{4(N_1^2\sigma_0^2+N_0^2\sigma_1^2)}{(N_0+N_1)^4}}
\label{eqn:qucounts1}
\end{equation}
\begin{equation}
\centering
u = \frac{N_{3} - N_{2}}{N_{3} + N_{2}},\, \sigma_u = \sqrt{\frac{4(N_3^2\sigma_2^2+N_2^2\sigma_3^2)}{(N_2+N_3)^4}}
\label{eqn:qucounts}
\end{equation}
where $N_i$ is the number of photons in the $i^{th}$ spot and $\sigma_i$ is the uncertainty that results from 
the photon noise. Therefore, overlapping of spots causes an 
artificially large difference in intensity of one pair of spots belonging to each affected star. The two
point spread functions (PSFs) cannot be de-blended, since the pipeline performs aperture photometry.
Typically, this contamination results in erroneously large degrees of polarization 
(but not necessarily, this can vary based on the relative brightness of the sources involved) and,
most notably, regular EVPAs (0$^{\circ}$, $\pm$45$^{\circ}$, $\pm$90$^{\circ}$). 
This follows from the definition of the EVPA, equation (\ref{eqn:EVPA}). 
If one of the vertical images of the star is artificially brightened,
for example $N_1 \ll N_2$, then $|u| \ll |q| \Rightarrow \chi\to\pm 45^{\circ}$. 
Whereas if one of the horizontal images is affected by a nearby source, 
e.g. $N_2 \ll N_3$, then $|q| \ll |u| \Rightarrow \chi\to0^{\circ},\pm 90^{\circ}$.
Fig \ref{fig:fingers} (left) shows the EVPA versus fractional linear polarization of all sources 
with at least 2 measurements found in the Polaris Flare field (5172 in total). 
Measurements of $p > 20\%$ are clustered around regular EVPAs - the clear signature of nearby star contamination. 

We remove such sources from the analysis in the following way. If any of the four spots of a star suffers from
confusion with another spot then we flag it as nearby contaminated. This flag applies if a source exists
within 3$\times$FWHM of a star spot. In cases where the spots of two stars happen to fall exactly on each other 
and are identified as a single source we check if any spot is assigned to more than one star. 
The effect of removing contaminated sources from the final catalog can be seen in Fig. \ref{fig:fingers} (right). 
All but two measurements with $p$ $>$ 20$\%$ were caused by proximity to other sources.

\begin{figure}
\centering
\includegraphics[scale=1]{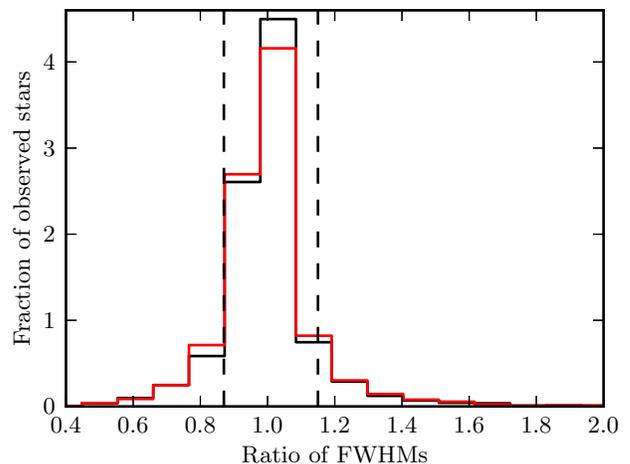}
\caption{Distributions of ratios of the FWHM between vertical (black) and horizontal (red) spots for a number of fields. Vertical lines mark the area that contains acceptable ratios.}
\label{fig:fwhm}
\end{figure}

Stars that are affected by reflections, and even other close-by stars in the case that the previous check fails,
can be removed by checking the ratio of the FWHM between two pairs of spots. The distribution of these ratios for
all stars found in all observed frames is shown in Fig. \ref{fig:fwhm}. 
We discard measurements lying outside the range 0.87 to 1.15 (vertical lines).

\subsection{Aperture optimization}

The RoboPol pipeline performs aperture photometry to measure the intensity (photon counts - N) of each spot. 
It then uses these values to calculate the Stokes parameters as shown in equations (\ref{eqn:qucounts1}) and 
(\ref{eqn:qucounts}). 
Photometry measurements are greatly affected by the choice of aperture size \citep[e.g.][]{Howell1989}. 
If the aperture is too large the value obtained suffers from background contamination and the signal to noise ratio 
is decreased. On the other hand, if the aperture is too small only a fraction of the total flux is measured. 
This is not a problem if the same fraction of photons are counted, since polarimetric measurements depend on 
the relative brightness of two spots. If, however, the PSFs of two spots belonging to a
source are different, then the fraction of the total flux measured is not the same and this introduces artificial 
polarization. 

A number of circumstances may affect the PSF of the four spots of a source.
Bad seeing or weather conditions (wind) sometimes cause sources to appear elongated instead of round. 
Also, the optical system of the instrument may distort the shape of the PSF and mainly the wings of the profile.
Typically, bright stars (whose wings are more prominent) are affected more severely than faint ones. Therefore, 
it is essential that photometry be performed with an aperture optimized for each source. Also, the complexity
of the optical system introduces some asymmetries in the PSFs of the vertical and horizontal images
of a star. Consequently, photometry must also be optimized for \emph{each} of the four images of a star.

We created a simple aperture optimization algorithm as an addition to the original pipeline, presented in
\cite{2014MNRAS.442.1706K}. 
Photon counts are measured within a circular aperture centred on each spot, 
while the background level is estimated within an outer concentric annulus that is separated from the 
inner aperture by a gap. 
The diameter of the background annulus is a constant multiple of the aperture size. 
The constant is different for faint and bright sources as the latter have more extended wings, 
so that the annulus does not contain any light from the source while retaining the smallest possible distance 
from the source for the background estimation.

By measuring the background subtracted photon flux within increasing apertures we create a growth curve for 
each spot. Each of the four growth curves of the source are fitted with a fourth degree polynomial, $P(x)$ 
(no errors are accounted for in the fit). 
The size of the aperture at which the normalized photon flux saturates is the optimum. 
To locate it in practice we look for the aperture size at which the rate of photon flux increase 
has reached some small value $\lambda$.
In other words, the optimal aperture is the root of the equation: 
\begin{equation}
\centering
\frac{dP}{dx}=\lambda P(x).
\label{eqn:growth-rate}
\end{equation} 
 
\begin{figure}
\centering
\includegraphics[scale=1]{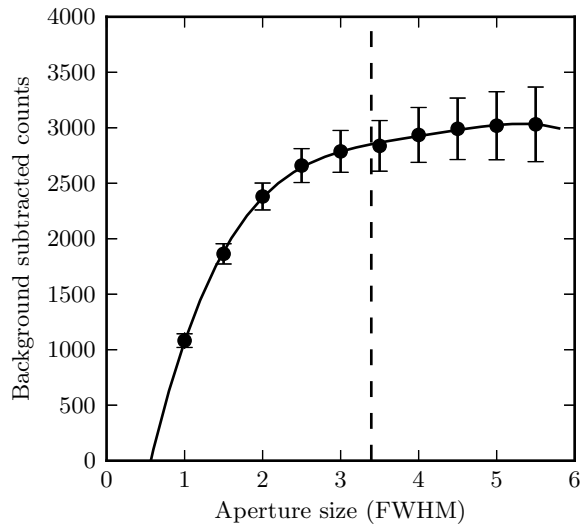}
\caption{Growth curve of one of the images of a star (circles show the number of background subtracted counts
for each aperture size). A fourth degree polynomial is fit to the
points. The optimal aperture is shown with the dashed line.}
\label{fig:polynomials}
\end{figure}

An example growth curve of one of the images of a star is shown
in Fig. \ref{fig:polynomials} (circles) along with its polynomial fit (solid line). The dashed vertical line 
shows the optimal aperture found by solving equation (\ref{eqn:growth-rate}). This aperture is used to measure 
the photon counts and noise ($N$, $\sigma$) of this spot.
The optimization is used for all data, including those collected for the instrument model calculation.

\subsubsection*{The choice of the value for $\lambda$}
\begin{figure}
\centering
\includegraphics[scale=1]{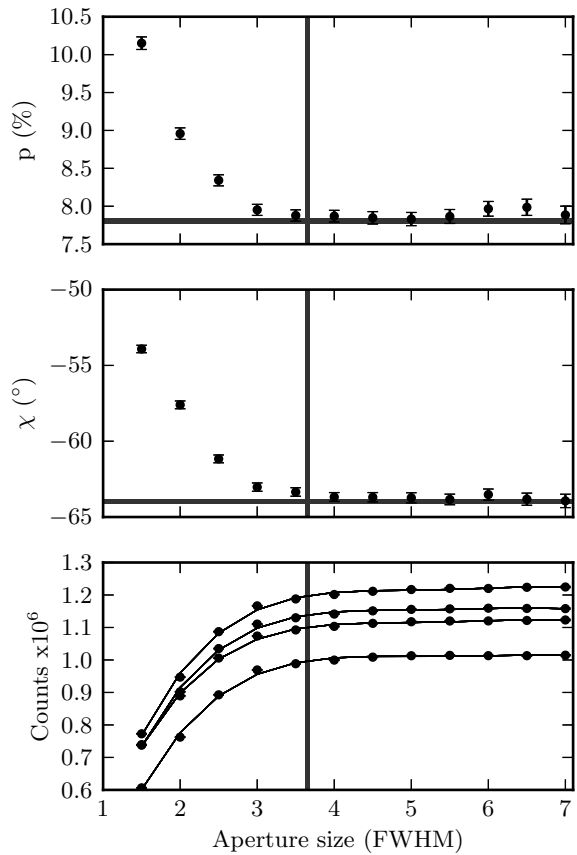}
\caption{Top: Fractional linear polarization of the standard star VI Cyg12 measured with different
 aperture sizes (multiples of FWHM). 
The horizontal gray band shows the literature value $\pm 1 \sigma$. 
The position of the vertical gray band shows the mean of the four optimal apertures
while its width represents their scatter.
Middle: EVPA measured with different aperture sizes. 
Bottom: Background subtracted number of counts with different aperture sizes (growth curve).
The growth curves for all four of images of the star are shown, along with a fourth degree polynomial fit.}
\label{fig:VCyg12_1-growth}
\end{figure}

To calibrate equation (\ref{eqn:growth-rate}) and determine the best value of $\lambda$, 
we created growth curves of polarization standard stars that were routinely observed in the field and 
measured their fractional linear polarizations and angles using all the different aperture sizes. 
Fig. \ref{fig:VCyg12_1-growth} shows the fractional linear polarization (top) and EVPA (middle) measured for 
VI Cyg12 with different apertures. As the aperture increases, these quantities saturate at some value consistent
with those found in the literature (gray bands). As aperture size continues to increase, the signal-to-noise
ratio worsens and also nearby sources may affect the measurement.
The point on the horizontal axis after which saturation occurs is the optimal aperture for this star.  
The parameter $\lambda$ was selected so that it reflects this transition. The bottom panel of 
Fig. \ref{fig:VCyg12_1-growth} shows the four growth curves of VI Cyg12 and the corresponding polynomial fits.
The vertical line shows the aperture that was chosen as optimal.

Because the standards observed with RoboPol are bright (typically 9-11 mag) we needed to extend this sample
to stars of lower brightness.
We selected 6 stars that were already observed in the field and observed them in the mask. 
We used the values found in the mask to define the optimal aperture for these sources when observed in the field. 
Finally, we optimized the parameter $\lambda$ so that it yields an accurate optimal aperture for 
most of the stars (both these 6 as well as the standards): $\lambda = 0.02$.

\subsection{Detection of dust specks using flat field images}

The design of RoboPol does not allow us to correct science images for irregularities in transmission and
uneven sensitivity throughout the field in conventional ways (e.g. by dividing pixel-by-pixel by a flat field
image). Because both sets of orthogonally polarized beams are projected on the same CCD, 
when recording an extended image (such as a flat) each point on the CCD is exposed to four rays tracing four
different optical paths through the instrument. 
In contrast, the photon counts we would like to correct (i.e. each of the four images
corresponding to a point source) arrive on the CCD through a single optical path since each beam 
corresponds to a different orientation of the plane of polarization. 
Light from the sky against which they are projected still arrives through four paths for each
pixel, but at different ratios, since the polarization of the sky differs between the moments of science and
flat field image acquisition. This makes ordinary flat fielding impossible.

\begin{figure}
\centering
\includegraphics[scale=1]{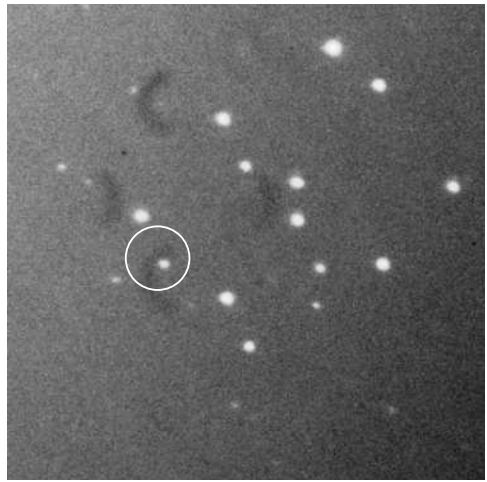}
\caption{Crescent dust patterns on the CCD. One of the four images of a star falls on a dust pattern (in the white circle).}
\label{fig:dust-spot}
\end{figure}

The global non-uniformity of the field (caused by vignetting) is corrected by the instrument model
as described in Section \ref{ssec:calibration}. Small-scale non-uniformities cannot be corrected for, but they can
be identified on flat-field images. Stars that happen to be affected by these small scale variations must be
excluded from the analysis. An example is shown in Fig. \ref{fig:dust-spot} where the crescent pattern produced 
by a dust speck is clearly visible in the exposure and coincides with one of the four spots of a star 
(circled in white).

We process flat-field images obtained in the evening and/or the morning of each night in the following way: we 
create a master flat by normalizing separate shots and taking the median. After that we fit a third degree 
polynomial to the master flat and subtract its value from each pixel, thus removing the large scale vignetting in the flat image.

At the position of each spot in the science image, we calculate the mode value ($F_{\rm bgr}$) and standard
deviation of counts ($\sigma_{\rm bgr}$) of pixels on the flat-field image 
that fall within an aperture with diameter equal to the background annulus.
In principle, by comparing these quantities on all four spots of a star we can determine whether any of them has
fallen on a dust speck, since this would cause significant variations in $F_{\rm bgr}$ and $\sigma_{\rm bgr}$.

To establish a set of reliable criteria that can identify most, if not all, dust-contaminated stars we analysed
data of the Be X-ray binary CepX4 \citep[e.g.][]{ulmer}, which is one of the most crowded fields observed with RoboPol \citep{2014MNRAS.445.4235R}.
We constructed a number of different quantities with the information from the flat field image. 
Those that proved most useful in revealing the effect of dust contamination were the following: 

\begin{itemize}
\item difference between the $\sigma_{\rm bgr}$ of a star's vertical (horizontal) spots 
($\Delta \sigma_{\rm bgr,v},\Delta \sigma_{\rm bgr,h}$),
\item difference between the background value of a star's vertical (horizontal) spots 
($\Delta F_{\rm bgr,v},\Delta F_{\rm bgr,h}$),
\item maximum $\sigma_{\rm bgr}$ (among four spots),
\item minimum $\sigma_{\rm bgr}$ (among four spots).
\end{itemize}

\begin{figure}
\centering
\includegraphics[scale=1]{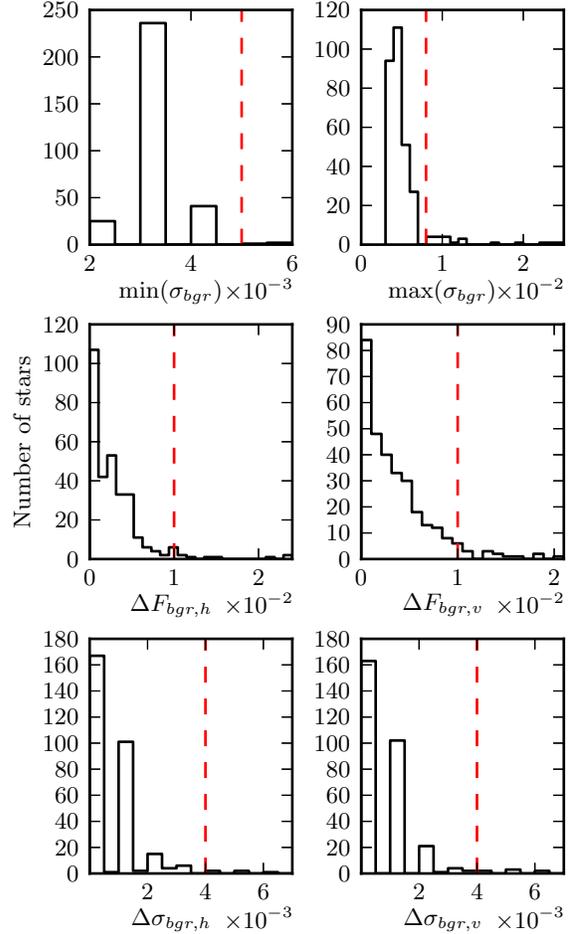}
\caption{Distributions of the values extracted from the flat fields and used for identifying dust specks. 
Values on the horizontal axis are measured in normalized counts.}
\label{fig:dust-distros}
\end{figure}

The distributions of all 6 quantities are shown in Fig. \ref{fig:dust-distros}. 
These quantities are measured in units of normalized counts in the processed master flat image.
The outliers of these distributions are stars that coincide with the most prominent dust specks.
According to these distributions we selected the thresholds depicted by vertical lines in Fig. 
\ref{fig:dust-distros}. 

Using these criteria we manage to eliminate only stars that are affected by the most obvious dust shadows.
A more sophisticated analysis is needed to identify more subtle anomalies on the CCD.

\subsection{Statistical assessment}
\label{sec:statisticcuts}

The standard observing strategy in the RoboPol project is to obtain multiple exposures of the same field.
We are thus able to use the stability of the measurements (in a statistical sense) to further reject stars 
with unreliable polarization measurements.
One reason for turning to a statistical treatment of the data is that even after the first stage of rigorous
cuts described in this section, some systematic errors are still present. 
These include faint dust specks, reflections from bright stars,
and in general, sources with properties around the various thresholds that were used. 

First, we choose to work with stars that have reliable measurements of the weighted mean of $p$ 
($\bar{p}/\bar{\sigma}_p \geq 2.5$).
The weighted mean is calculated by substituting into equation (\ref{eqn:P}) the weighted mean $q$ and $u$
values of a star.

\begin{figure}
\centering
\includegraphics[scale=1]{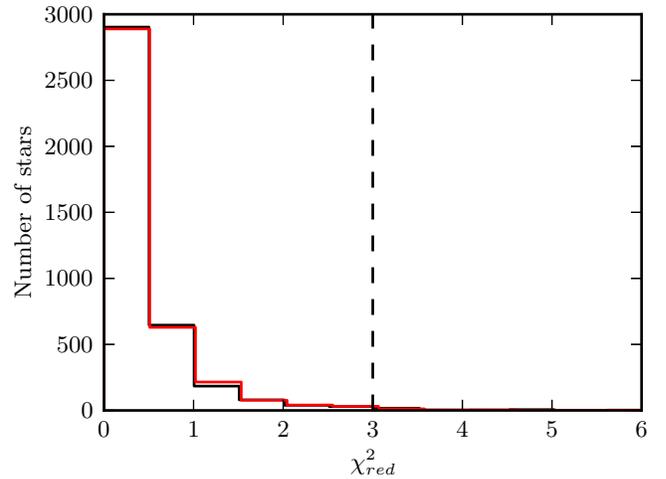}
\caption{Distributions of $\mathcal{X}^2_{red}$ of $q$ (black) and $u$ (red) of stars with $\bar{p}/\bar{\sigma}_p \geq 2.5$. The vertical line shows the selected threshold. }
\label{fig:rcs}
\end{figure}

One way to quantify the statistical significance of the differences between the $n$ measurements of a star 
is by computing the reduced $\mathcal{X}^2$ ($\mathcal{X}^2_{red}$) of all of its $q$ and $u$ measurements:
\begin{equation}
\centering
\mathcal{X}^2_{red,q} = \frac{1}{n-1} \sum_{j=1}^n \frac{(q_j - \bar q)^2}{(\sigma_{q,j})^2}
\end{equation}
and similarly for $\mathcal{X}^2_{red,u}$.
By placing a threshold in the value $\mathcal{X}^2_{red}$ we can eliminate stars that deviate 
from the expected normal behavior. 
The threshold was selected so as to remove the tail of the distribution of all $\mathcal{X}^2_{red}$
values of stars in the Polaris Flare region.
The distributions of these values for $q$ and $u$ measurements are shown in Fig. \ref{fig:rcs} as well as the selected threshold (vertical line). 

Stars that still remain after these cuts and show signs of some type of contamination visible by eye
on the raw science images were removed by hand. These include types of contamination already presented
in this section as well as projected double stars, for which the analysis does not account.

\section{Results and Discussion}
\label{sec:PF}
The analysis provides us with 648 stars with reliable $p$ and $\chi$ measurements. 
They are presented in the online table accompanying this paper (Table \ref{tab:data}). 

\begin{figure}
\centering
\includegraphics[scale=1]{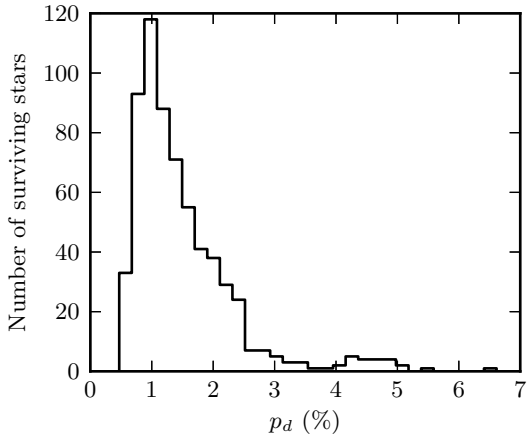}
\caption{Distribution of debiased fractional linear polarizations of all 648 sources resulting from the analysis.}
\label{fig:pdebiased}
\end{figure}

\begin{figure}
\centering
\includegraphics[scale=1]{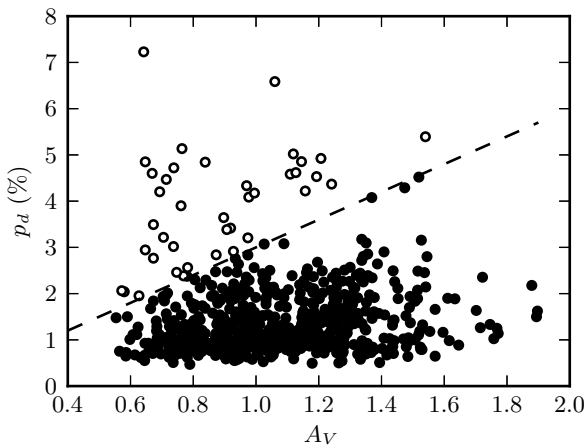}
\caption{Debiased polarization percentage vs. visual extinction, $A_V$ for all our 
reliably measured stars. The dashed line shows the maximum observable $p$ at all $A_V$ ($p$ = $0.03
A_V$). Stars above the black line are marked with empty circles.}
\label{fig:pav}
\end{figure}

The distribution of debiased fractional linear polarizations of all these sources is shown in 
Fig. \ref{fig:pdebiased}. The median of the distribution is at 1.3\%.  
Fig. \ref{fig:pav} shows the debiased polarization percentage
against visual extinction derived by the SDSS colors as in \cite{schlafly}. 
The dashed line shows the empirically determined upper limit in polarization at a given $A_V$:  
$p/A_V = 0.03$ \citep{serkowski}.
We mark sources above this limit with empty circles and use this line as a threshold. 
Sources above the line are considered separately as their polarizations may have an intrinsic contribution.

\begin{figure*}
\centering
\includegraphics[scale=1]{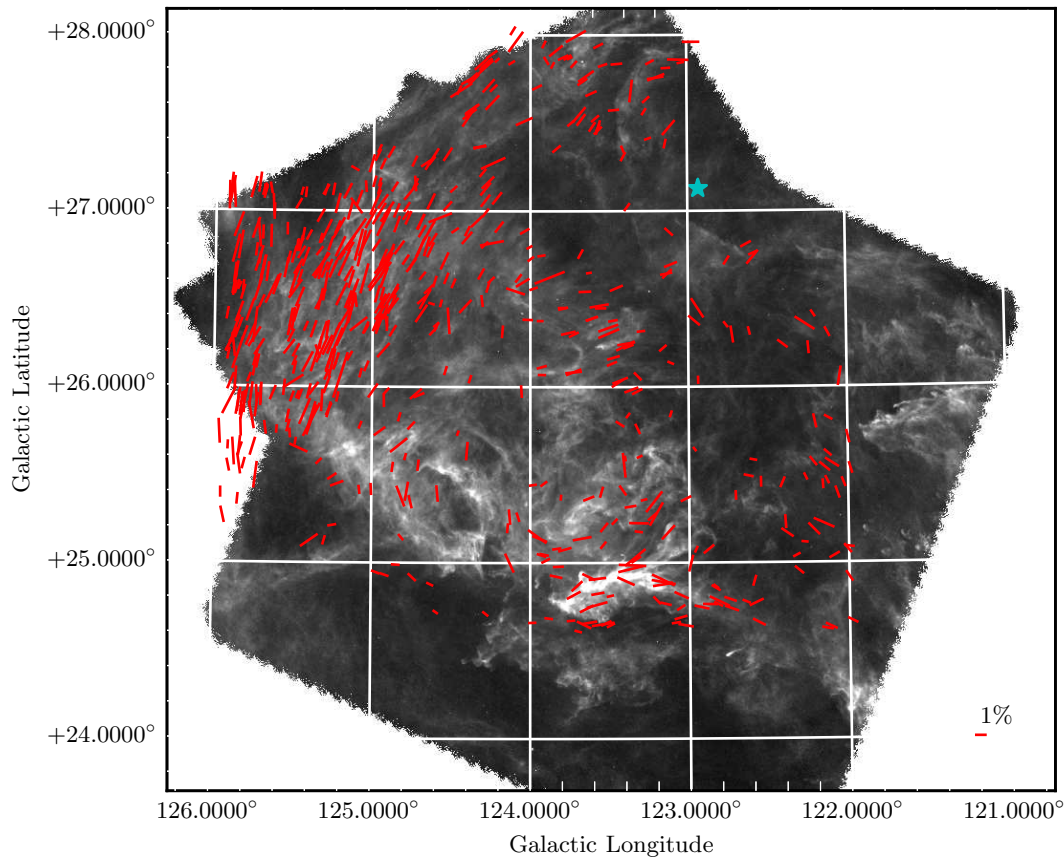}
\caption{Polarization segments over plotted on top of the Herschel 250 $\mathrm{ \mu m}$ image of the Polaris Flare. The length of each segment is proportional to the debiased $p$ of the star. The horizontal segment at l = 121$^\circ$ is for scale. The blue star marks the position of the North Celestial Pole.}
\label{fig:map}
\end{figure*}

In order to construct the polarization map of the region we transform all EVPAs 
(measured with respect to the North-South celestial pole direction) into galactic angles according to
\cite{stephens}\footnote{see erratum published in MNRAS.}.
We plot the polarization segments of all stars below the $p_d-A_V$ line of Fig. \ref{fig:pav} at each star 
position on the Herschel 250 $\mathrm{\mu m}$ image of the Polaris Flare \citep{miville} 
in Fig. \ref{fig:map}. The length of each segment is proportional to the debiased $p$ of the star, calculated
using equation (\ref{eqn:debiasing}).
The most striking feature of the polarization map is the extended ordered pattern at large longitudes.
In this region the plane-of-the-sky magnetic field appears to be oriented in approximately the same direction
as that of the faint striations seen in dust emission. A very similar pattern has been observed in the much 
denser Taurus Molecular Cloud \citep{chapman}.  
Segments at the largest longitudes are mostly parallel to lines of constant longitude, 
following the projected curvature of a vertical cloud structure that is partly cut-off by the map edges.
A border appears to exist, spanning the diagonal of region (124$^\circ$, 125$^\circ$), (26$^\circ$, 27$^\circ$). 
Segments below this virtual line form a loop, or eddy-like feature centred at 
(124$^\circ$, 25.5$^\circ$) that covers latitudes down to 24.5$^\circ$ and longitudes down to 123$^\circ$. 
In the south, segments that are projected on the dense filamentary region, also known as the MCLD 123.5+24.9 cloud,
appear to be parallel to the axis of the filament and its surrounding less dense gas. 
A detailed quantitative comparison of the magnetic field as revealed by the map with the dust column density from
Herschel will be presented in a follow-up paper.

The general structure of the plane-of-the-sky magnetic field in this cloud agrees qualitatively with that inferred
from the polarized emission seen by the Planck satellite \citep{planckXX}. Even though the resolution of
the presented map does not allow for a detailed comparison, the orientation of the ordered east part is in fair agreement with that seen in our map. Also, the central-southwest part in the Planck map does show a discontinuity
of the projected field orientation that could be a sign of the loop that we observe. 

The proximity of the cloud suggests that the level of
contamination by dust foreground to the cloud is insignificant. Stars lying in front of the cloud will most
likely exhibit very low polarization ($\ll 1\%$) and so would not comply with the $p/\sigma_p$ threshold,
thus they would not affect the map.

\begin{figure}
\centering
\includegraphics[scale=1]{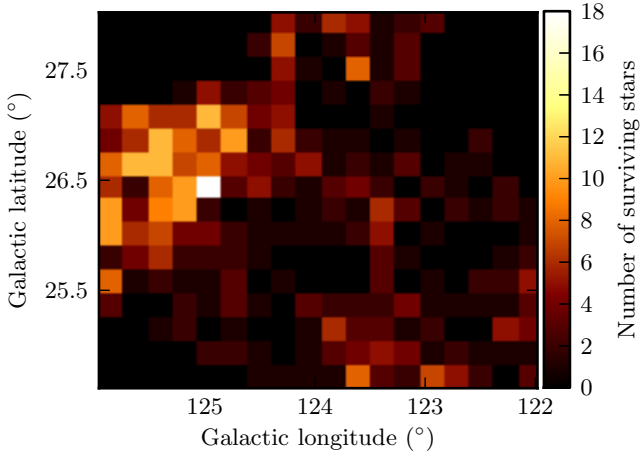}
\caption{Number of stars in the map per field across the sky. The size of the bins corresponds to that of the
field of view. The non-uniformity is a result of the $p/\sigma_p$ cut.}
\label{fig:stardist}
\end{figure}

\begin{figure}
\centering
\includegraphics[scale=1]{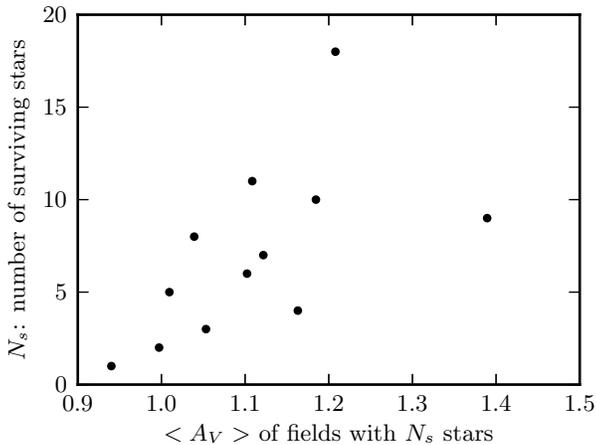}
\caption{Number of stars ($N_s$) in Fig. \ref{fig:map} versus mean $A_V$ in all fields with $N_s$.}
\label{fig:n-av-corr}
\end{figure}

The distribution of stars for which we have reliable polarization measurements is not uniform. 
Segments at higher galactic latitude and longitude are denser than at the lower part of the map.
Fig. \ref{fig:stardist} shows the number of stars in the map binned across the entire observed region. 
The bin size corresponds to that of the field of view. 
The brighter regions (containing more stars per bin) are in the area with ordered plane-of-the-sky 
magnetic field. This non-uniformity is not due to variations in the stellar density across the observed region. 
It appears as a result of the $p/\sigma_p$ cut. We find no correlation between this pattern and
observing conditions (i.e. seeing, elevation, moon phase). 
For all fields with a given number of surviving stars ($N_s$), we calculate the mean 
extinction $\langle A_V\rangle$. There is a clear correlation between the two, as can be seen in 
Fig. \ref{fig:n-av-corr}.
We find that the Pearson correlation coefficient between these two sets is 0.59.
We therefore conclude that this effect is most likely not the result of some systematic error,
but that of the cloud properties. A possibility that could give rise to this effect is a magnetic field
whose direction changes from mostly on the plane-of-the-sky in the upper left part of the map, to 
having a more pronounced component along the line of sight towards the lower right.
 
\subsection*{Potentially intrinsically polarized sources}
\begin{figure}
\centering
\includegraphics[scale=1]{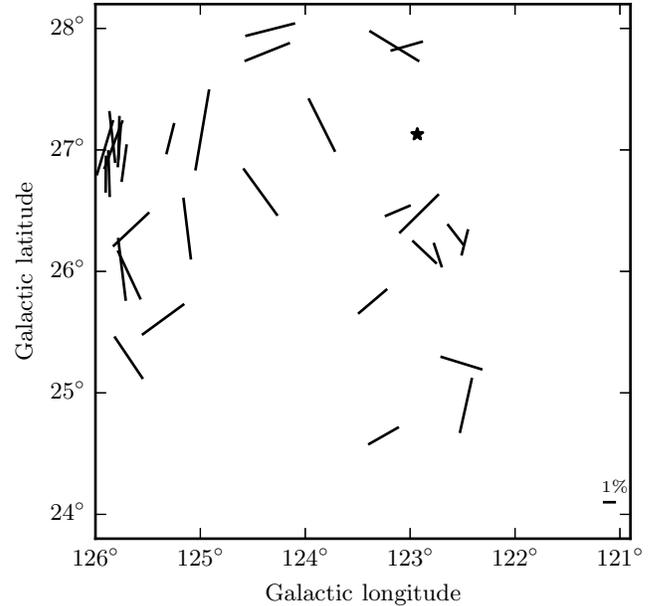}
\caption{Polarization segments of stars above the P-$A_V$ line shown in Fig. \ref{fig:pav}. 
The black star shows the position of the North Celestial Pole while the horizontal segment in the bottom
right sets the scale (1\%).}
\label{fig:abovepav}
\end{figure}

We plot the polarization segments of sources above the dashed line of Fig. \ref{fig:pav} 
separately in Fig. \ref{fig:abovepav}
to be easily distinguished from those whose polarization is primarily affected by the magnetic field of 
the cloud. The length of the polarization segment of each star is proportional to its debiased $p$.
The orientations of some segments are correlated with the general direction of the plane-of-the-sky magnetic 
field map of Fig. \ref{fig:map}. This is not surprising since the $p$-$A_V$ line is empirical. 
Therefore our choice of setting a threshold based on that line is conservative.

We investigate the possibility that a number of the 39 sources falling above the $p_d - A_V$ line in 
Fig. \ref{fig:pav} could be quasar candidates. 
Multi-wavelength data in this region are sparse, so cross-correlations with our sample were not particularly
fertile. The low resolution of radio data renders direct identification of optical counterparts impractical. 
For most highly polarized sources (over the $p_d - A_V$ line) we only managed to find data
from the USNOB and 2MASS catalogs. \cite{2mass} presented the color properties of quasar and AGN candidates in the 
2MASS catalog. They demonstrated that candidates can be found preferentially at certain regions of 
color-color diagrams. Only one of our sources seems to marginally fit into this category. 
It should be noted, though, that these values have not been redshift-corrected.

\section{Summary}
\label{sec:summary}
We have presented optical linear polarization measurements of stars projected on the Polaris Flare field.
These measurements reveal the plane-of-the-sky magnetic field structure of the cloud. 
The observations span about 10 $\rm deg^2$ of the region and have been conducted with the RoboPol polarimeter
in the R-band. We presented adjustments to the automated data reduction pipeline that were necessary for the
analysis of sources in the entire $13'\times13'$ field of view. We have investigated possible sources of systematic 
errors and have presented our methods for correcting for each one.

We have produced a map of 648 polarization segments showing the magnetic field structure of the cloud as projected
on the plane of the sky. The median debiased $p$ is 1.3\%. The projected field shows a complicated,  
ordered structure throughout the map. 
To the top left part of the map, the field is aligned with the striations seen in dust emission.
The bottom right parts show the presence of an eddy-like feature spanning roughly 2 degrees in diameter.
Our results compare well with the Planck map of polarized emission of the cloud.
The distribution of stars with reliable polarization measurements across the field is not uniform, with 
most stars lying in the top left of the region. This is most likely due to the intrinsic properties of the 
magnetic field structure.

\begin{table*}
\centering
\caption{Reliable polarization measurements in the Polaris Flare region (full table online).}
\begin{tabular}{|c|c|c|c|c|c|c|c|c|}
\hline
R.A. & Dec & $l (^\circ)$& $b (^\circ)$ & $p_d$ (\%) & $\sigma_p$ (\%) & $\chi (^\circ)$ & $\sigma_\chi (^\circ)$& $\theta_{gal} (^\circ)$\\ 
\hline
\hline
53.78514 & 87.71787 & 124.58667 & 25.39440 & 1.4 & 0.5 & 44 & 7 & 4 \\ 
67.42559 & 88.23095 & 124.53694 & 26.09361 & 0.8 & 0.3 & 36 & 8 & 162 \\ 
66.70140 & 88.02690 & 124.70368 & 25.95310 & 1.0 & 0.3 & -14 & 8 & 112 \\ 
75.26493 & 88.39611 & 124.51862 & 26.37650 & 2.3 & 0.6 & 65 & 7 & 3 \\ 
\end{tabular}
\label{tab:data}
\end{table*}

\section*{Acknowledgements}

We thank A. Kougentakis, G. Paterakis, and A. Steiakaki,
the technical team of the Skinakas Observatory.  
The University of Crete group acknowledges support by the ``RoboPol'' project,
implemented under the ``ARISTEIA'' Action of the ``OPERATIONAL PROGRAMME EDUCATION AND
LIFELONG LEARNING'' and is co-funded by the European Social Fund (ESF) and Greek National
Resources. The Nicolaus Copernicus University group acknowledges support from the Polish 
National Science Centre (PNSC), grant number 2011/01/B/ST9/04618. 
This research is supported in part by NASA grants NNX11A043G and NSF grant AST-1109911. 
V.P. acknowledges support by the European Commission Seventh Framework Programme (FP7) 
through the Marie Curie Career Integration Grant PCIG10-GA-2011-304001 ``JetPop''. 
K.T. acknowledges support by FP7 through the Marie Curie Career Integration Grant 
PCIG-GA-2011-293531 ``SFOnset''. V.P., E.A., I.M., K.T.,
and J.A.Z. would like to acknowledge partial support from the EU FP7 Grant
PIRSES-GA-2012-31578 ``EuroCal''.  I.M. is supported for this research through a stipend
from the International Max Planck Research School (IMPRS) for Astronomy and Astrophysics
at the Universities of Bonn and Cologne. M.B. acknowledges support from the International
Fulbright Science and Technology Award. T.H. was supported by the Academy of
Finland project number 267324. The RoboPol collaboration acknowledges observations support
from the Skinakas Observatory, operated jointly by the University of Crete and the Foundation for
Research and Technology - Hellas.  Support from MPIfR, PNSC, the Caltech Optical
Observatories, and IUCAA for the design and construction of the RoboPol polarimeter is
also acknowledged.

This research has used data from the NASA/IPAC Extragalactic Database (NED) which is operated by the Jet Propulsion Laboratory, California Institute of Technology, under contract with the National Aeronautics and Space Administration. This research has made use of the SIMBAD database, operated at CDS, Strasbourg, France as well as the VizieR catalogue access tool, CDS, Strasbourg, France. 
This work made use of APLpy, an open-source plotting package for Python hosted at http://aplpy.github.com, Astropy, a community-developed core Python package for Astronomy \citep{2013A&A...558A..33A}, matplotlib, a Python library for publication quality graphics \citep{matplotlib} and the Python library SciPy \citep{scipy}.

\end{document}